\documentclass[showpacs,preprintnumbers,amsmath,amssymb,prb,twocolumn,floatfix,amssymb]{revtex4}
\usepackage[dvips]{graphicx,color}
\usepackage{dcolumn}
\usepackage{amsmath,amssymb,bbold,bm}

\begin{document}
\title{Excitons and Optical Absorption on the Surface of a Strong Topological Insulator with a Magnetic Energy Gap}
\author{Ion Garate$^{1,2}$ and M. Franz$^{1}$}
\affiliation{$^1$Department of Physics and Astronomy, The University of British Columbia, Vancouver, BC V6T 1Z1, Canada}
\affiliation{$^2$Canadian Institute for Advanced Research, Toronto, ON M5G 1Z8, Canada.}
\date{\today}
\begin{abstract}
We present a theoretical study of interacting electron-hole pairs located on a magnetized surface of a strong topological insulator (TI).
The excitonic energy levels and the optical absorption on such surface display unique and potentially measurable features such as (i) an enhanced binding energy for excitons whose total angular momentum is aligned with the magnetic exchange field, (ii) a stark dependence of the optical absorption on the direction of the magnetic exchange field as well as on the chirality of the incident light,  and (iii) a tunable center-of-mass motion of spinful excitons induced by particle-hole asymmetry in the exchange field.
Our predictions are relevant to surfaces of magnetically doped TIs or surfaces coated with magnetic films, in addition to TI nanowires placed under longitudinal magnetic fields. 
\end{abstract}
\maketitle

\section{Introduction}

The discovery of strong topological insulators has culminated in a new classification scheme for solids.\cite{hasan2010} 
Strong TIs are bulk insulators endowed with a topological invariant that manifests itself through robust metallic surface states containing a nontrivial Berry phase.
These surface states have inspired numerous attractive proposals for spintronics applications\cite{ts} because they are strongly spin-orbit coupled and because they exhibit essential robusteness against non-magnetic perturbations.  In addition, they have a linear energy-momentum dispersion at low energies, they may display a half-quantized quantum Hall effect when time-reversal symmetry is broken, and they may host exotic particles such as Majorana fermions when placed in proximity to a superconductor.

Thus far both experimental and theoretical studies of topological surface states have either ignored electron-electron interactions or treated them ad hoc through mean field\cite{santos2010} theories.
This omission is partly justified by the fact that weak, short-range interactions between massless Dirac fermions are irrelevant in the renormalization group (RG) sense. 
Experimental support for the RG argument originates from the sizeable Fermi velocities ($v_F\simeq 5\times 10^5 {\rm m/s}$) and the large static dielectric constants ($\epsilon\sim 50-100$) found in several well-established TIs, which considerably reduce the value of the bare Coulomb coupling $g\equiv (e^2/\epsilon)/(2\hbar v_F)$ ($g\lesssim 0.05$).
However, there appears to be no fundamental principle that forbids the existence of TIs with smaller\cite{diel} dielectric constants and thereby more strongly interacting surface states.
This expectation is made more plausible by the ongoing and rapid expansion of experimentally confirmed TIs, which currently encompass a variety of bismuth compounds \cite{hasan2010} as well as chalcogenides\cite{sato2010}.
For instance TlBiSe$_2$, recently\cite{sato2010} identified as a TI,  has been previously reported\cite{mitsas1992} to display a relatively small static dielectric constant of $\epsilon\simeq 25$ in thin films.
Likewise HgTe, which has been recently\cite{brune2011} coaxed into a strong topological insulating state, has $\epsilon\simeq 20$ in bulk. 
Moreover, the irrelevance under RG of electron-electron interactions is questionable when the chemical potential of the topological surface states is located inside a magnetically generated energy gap (which introduces an infrared cutoff in the RG flow equations).

This paper aims to initiate a systematic study of the interplay between spin-orbit interactions, broken time reversal symmetry, and Coulomb interactions on surfaces of strong TIs.
In particular, we focus on low-temperature optical properties of topological surface states whose chemical potential lies inside a magnetically generated energy gap.
The underlying microscopic mechanism for this energy gap at the surface is an exchange field in the direction perpendicular to the surface, which can be produced by magnetic doping\cite{chen2010} or else by deposition of a ferromagnetic insulator with perpendicular anisotropy (the latter is yet to be realized experimentally). 
Of course, a magnetic gap may also be opened by the application of a perpendicular magnetic field; however, in this case the surface states are modified qualitatively owing to the formation of Landau levels.
Herein we disregard orbital effects\cite{tse} and thus limit ourselves to exchange-induced magnetic gaps.
 
The simplest interacting problem one can conceive in this playground is that of a surface electron-hole excitation across the magnetic bandgap.
Some of the peculiarities of this two-body problem are made apparent in Section II.
On one hand, the relative motion of the electron-hole pair is inextricably linked to its center-of-mass motion, which is conducive to phenomena like the spin-Hall effect of excitons.
On the other hand, optically generated electron-hole pairs have a nonzero center-of-mass group velocity provided that the exchange field is particle-hole asymmetric.
While the former property is also present in certain ordinary semiconductors,\cite{shex} the latter trait is exclusive of topological surface states.

In Section III we consider a two dimensional surface with translational and rotational symmetry.
In this case, broken time reversal symmetry and spin-orbit coupling conspire to produce Coulomb interaction matrix elements that favor a nonzero total angular momentum for the electron-hole pairs.
This skeweness of the Coulomb interaction is transferred to the binding energy of dilute excitons, which becomes largest for the p-wave angular momentum channel.
When the surface is irradiated with linearly polarized light, this p-wave exciton is optically bright and appears prominently in the absorption spectrum of systems with $g>0.05$.
In contrast, for circularly polarized light the dominant p-wave exciton is optically dark or bright depending on the handedness of the light as well as on the orientation of the exchange field.

In Section IV we turn to low dimensional TIs. 
Typically nanowires show more pronounced excitonic and electric-field effects in their optical absorption spectrum. 
We verify that, in presence of particle-hole asymmetry, optically generated excitons have a nonzero center-of-mass group velocity.
Incidentally, we establish the appearance of electric-field-induced Franz-Keldysh oscillations.
Nanostructures made of TIs are being routinely fabricated,\cite{kong2010} which may motivate optical measurements that test the effects discussed herein.

Our quantitative results contain unknown parameters such as the surface dielectric constant and the lifetime of the excitons; it may be possible to extract these from photoluminescence experiments that are presently underway\cite{laforge2010} in TIs.

\section{General considerations}

In this section we consider an interacting electron-hole pair located at the surface of a magnetized STI, and discuss some of its generic aspects.
The Hamiltonian describing this problem is
\begin{eqnarray}
\label{eq:h1}
{\cal H}&=&v_F{\boldsymbol\sigma}_e\cdot(\hat{z}\times {\bf p}_e)+\Delta_e \hat\Omega\cdot {\boldsymbol\sigma}_e\nonumber\\
&-&v_F{\boldsymbol\sigma}_h\cdot(\hat{z}\times {\bf p}_h)+\Delta_h \hat\Omega\cdot {\boldsymbol\sigma}_h-\frac{e^2}{\epsilon r}-e {\bf {\cal E}}\cdot{\bf r},
\end{eqnarray}
where $v_F$ is the Fermi velocity, $\hat z$ is the unit vector normal to the surface, ${\bf p}_{e(h)}=-i\partial/\partial {\bf r}_{e(h)}$ is the momentum of the electron (hole), ${\bf r}_{e(h)}$ is the position of the electron (hole), ${\bf r}={\bf r}_e-{\bf r}_h$ is the relative coordinate, ${\boldsymbol\sigma}_e={\boldsymbol\sigma}\otimes {\bf 1}$ is the spin operator acting on the electron (${\boldsymbol\sigma}$ is a vector of Pauli matrices and ${\bf 1}$ is the $2\times 2$ identity matrix), and ${\boldsymbol\sigma}_h={\bf 1}\otimes{\boldsymbol\sigma}$ is the spin operator acting on the hole. 
${\bf \Delta}_{e(h)}=\Delta_{e(h)}\hat \Omega$ is the exchange field sensed by the electron (hole), which can be generated either by magnetically ordered impurities doped in the STI bulk, sprinkled on the surface\cite{chen2010} or else by deposition of a ferromagnetic insulator overlay.
$\hat\Omega$ denotes the direction of the magnetization and is assumed to be spatially uniform.
The exchange field originates from virtual transitions of the surface fermions onto localized magnetic ions and need not be the same for electrons and holes (i.e. $\Delta_e\neq\Delta_h$).
The reasons for this asymmetry are twofold.
On one hand, the exchange (or Kondo) coupling between itinerant fermions and localized magnetic moments is generally energy-dependent,\cite{hewson} i.e. $\Delta=\Delta(\epsilon)$ where $\epsilon=\hbar v_F k$.
On the other hand, the magnetic impurities are generally not particle-hole symmetric, i.e. adding an electron to the local moment costs different energy from subtracting one. 
It follows that $\lim_{\epsilon\to 0}\partial\Delta/\partial\epsilon\neq 0$.
In Eq.~(\ref{eq:h1}) we adopted $\Delta(\epsilon)=\Delta_e\Theta(\epsilon)+\Delta_h\Theta(-\epsilon)$ (where $\Theta(x)$ is the step function), which constitutes the simplest realization of broken particle-hole symmetry in the exchange coupling.
While often ignored for single-particle physical observables, $\Delta_e\neq\Delta_h$ may lead to potentially observable two-body effects as indicated below.
Throughout this work we assume $\Omega_z\neq 0$, which creates an energy gap of $(\Delta_e+\Delta_h)\Omega_z$ for non-interacting particle-hole excitations.
In addition, we assume that the Fermi energy of the surface states is located within this energy gap.
The attractive Coulomb interaction between the electron and the hole is screened by the static dielectric constant $\epsilon$, whose value is expected to be a weighted average of the dielectric constants corresponding to the bulk TI and the vacuum (or the ferromagnetic insulator).
The larger weight is ascribed to the medium into which the surface states decay more slowly. 
${\bf {\cal E}}$ is the in-plane electric field acting on the surface states.
 
Adopting center-of-mass (COM) and relative coordinates, Eq.~(\ref{eq:h1}) can be rewritten as 
\begin{eqnarray}
\label{eq:h2}
{\cal H}&=& v_F\left({\boldsymbol\sigma_e}-{\boldsymbol\sigma_h}\right)\cdot\left(\hat{z}\times{\bf P}+\frac{\Delta_-}{v_F}\hat{\Omega}\right)\\
&+&\frac{v_F}{2}\left({\boldsymbol\sigma_e}+{\boldsymbol\sigma_h}\right)\cdot\left(\hat{z}\times{\bf p}+\frac{\Delta_+}{v_F}\hat{\Omega}\right)-\frac{e^2}{\epsilon r}-e {\bf {\cal E}}\cdot{\bf r},\nonumber
\end{eqnarray}
where ${\bf p}={\bf p}_e-{\bf p}_h$ is the relative momentum, ${\bf P}=({\bf p}_e+{\bf p}_h)/2$ is the COM momentum and $\Delta_\pm=(\Delta_e\pm\Delta_h)/2$.
${\boldsymbol\sigma_e}+{\boldsymbol\sigma_h}$ is the total spin operator for the electron-hole pair, and it couples to the relative momentum.
This can be easily understood by recalling that, for a single Dirac fermion, the spin of an electron in the ``conduction'' band with momentum ${\bf k}$ is perfectly antialigned with the spin of an {\rm electron} with the same momentum in the ``valence'' band.
The in-plane component of the exchange field can be absorbed by a shift in the relative momentum of the electron-hole pair, and therefore it is inconsequential. 

Because both electric fields and Coulomb interactions couple only to the relative coordinate, the COM momentum ${\bf P}$ is a good quantum number.
Within the envelope function approximation, the conservation of ${\bf P}$ enables a factorization of the electron-hole wavefunction via $\Psi({\bf R},{\bf r})=\exp(i {\bf P}\cdot{\bf R}) \chi({\bf P},{\bf r})$, where ${\bf R}$ is the COM coordinate and $\chi$ is the ``relative'' wave function.
The co-dependence of $\chi$ on ${\bf P}$ and ${\bf r}$ hints at the non-separability between the center-of-mass motion and the relative motion.
In effect, the COM and relative parts in Eq.~(\ref{eq:h2}) (i.e. the first and second lines) do not commute with one another.
This observation applies to graphene and carbon nanotubes as well;\cite{sabio2010} however, unlike in graphene and carbon nanotubes, in TIs momentum is coupled with spin rather than with a (less manipulable) sublattice degree of freedom.
The coupling between COM and relative degrees of freedom creates prospects for spintronics applications, such as a spin-Hall effect of excitons \cite{shex} derived from the Berry connection ${\cal A}_j=i\int d{\bf r}\chi^*\partial\chi/\partial P_j$. 

Aside from ${\bf P}$, the $z$-component of the electron-hole pair's total angular momentum is also a good quantum number so long as $P={\cal E}=\Delta_-=0$.
Indeed one can readily verify that $\lim_{P,{\cal E},\Delta_-\to 0} {\cal R}^{-1}{\cal H} {\cal R}={\cal H}$ for ${\cal R}=\exp(i s^z \phi) \exp(i l^z \phi)$, where $s^z=\sigma_e^z+\sigma_h^z$, $l^z=x p_y- y p_x$ and $\phi$ is an arbitrary angle of rotation.
In Section III we shall take advantage of this  U(1) symmetry in order to simplify the calculation of the excitation spectrum. 

An intriguing peculiarity of topological surface states is that particle-hole asymmetry in the exchange field produces a fictitious electromagnetic vector potential ${\bf a}=\Delta_-/v_F\hat{z}\times\hat\Omega$ for the center-of-mass coordinate. 
Therefore, the mechanical or physical COM momentum of the electron-hole pair, ${\boldsymbol \Pi}={\bf P}-{\bf a}$, is different from the canonical momentum ${\bf P}=-i\hbar\partial_{\bf R}$. 
Accordingly, optically excited electron-hole pairs (which are essentially vertical across the bandgap and therefore have ${\bf P}=0$) contain a nonzero mechanical momentum.
As we show in Section IV, the group velocity of these excitons is given by $v_{\bf g}\simeq v_F(\Delta_-/\Omega_z \Delta_+) \hat{z}\times\hat{\Omega}$ for $v_F |{\bf a}|\ll\Omega_z\Delta_+$ (which implies $v_g<<v_F$), and consequently their motion may be controlled by tilting the direction of magnetization away from $\hat z$.
Admittedly $\Delta_-$ itself is not easily tunable in a single topological surface. 
However, nanoribbons with two Dirac cones (one on each surface) offer a platform wherein $\Delta_-$ is tunable (by changing the relative magnetization of the two surfaces) for spatially indirect electron-hole pairs.
In this paper we limit ourselves to topological surfaces with a single Dirac cone. 


\section{Electron-hole pairs on a two-dimensional surface}

This section is devoted to the characterization of optically generated excitons and their photoluminescence on a two dimensional topological surface with a magnetic energy gap.
For simplicity we assume $\Delta_e=\Delta_h=\Delta$ and ${\bf {\cal E}}=0$.
Moreover we take ${\bf\Delta}=\Delta\hat{z}$, which involves no loss of generality because the in-plane component of the exchange field can be shifted away by redefining a new origin in momentum space.
These choices render the problem more easily soluble in momentum space, where the low energy effective Hamiltonian may be written as ${\cal H}={\cal H}_0+{\cal U}$, with
\begin{eqnarray}
{\cal H}_0 &=& \sum_{\alpha=c,v}\sum_{{\bf k}} E_{\alpha,{\bf k}} \Psi^\dagger_{\alpha,{\bf k}} \Psi_{\alpha,{\bf k}}\nonumber\\
{\cal U} &=&\frac{1}{L^2}\sum_\alpha \sum_{{\bf k,k',q}}\frac{1}{2} V^{\rm (A)}_\alpha \Psi^\dagger_{\alpha,{\bf k+q}}\Psi^\dagger_{\alpha,{\bf k'-q}} \Psi_{\alpha,{\bf k'}} \Psi_{\alpha,{\bf k}}\nonumber\\
&+&\frac{1}{L^2}\sum_{{\bf k,k',q}} V^{\rm (E)} \Psi^\dagger_{c,{\bf k+q}}\Psi^\dagger_{v,{\bf k'-q}} \Psi_{v,{\bf k'}} \Psi_{c,{\bf k}},
\end{eqnarray}
$E_{{\bf k},c}=-E_{{\bf k},v}\equiv E_k=\sqrt{\hbar^2 v_F^2 k^2 +\Delta^2}$ is the energy dispersion for non-interacting massive Dirac fermions (mass $\propto\Delta$), and
\begin{eqnarray}
\label{eq:interact}
V^{\rm (A)}_\alpha &=&\langle \alpha,{\bf k+q}; \alpha, {\bf k'-q}|{\cal U}|\alpha,{\bf k};\alpha,{\bf k'}\rangle \nonumber\\ 
V^{\rm (E)} &=& \langle c,{\bf k+q}; v, {\bf k'-q}|{\cal U}|c,{\bf k};v,{\bf k'}\rangle
\end{eqnarray}
are the intraband and interband Coulomb matrix elements.
The operators that create Bloch eigenstates in the ``conduction'' and ``valence'' bands (hereafter denoted as $|c{\bf k}\rangle$ and $|v {\bf k}\rangle$) are $\Psi^\dagger_{c,{\bf k}}=\cos(\theta_{\bf k}/2) c^\dagger_{{\bf k},\uparrow}+\exp(i\varphi_{\bf k})\sin(\theta_{\bf k}/2) c^\dagger_{{\bf k},\downarrow}$ and $\Psi^\dagger_{v,{\bf k}}=\sin(\theta_{\bf k}/2) c^\dagger_{{\bf k},\uparrow}-\exp(i\varphi_{\bf k})\cos(\theta_{\bf k}/2) c^\dagger_{{\bf k},\downarrow}$, respectively. 
$c^\dagger_{{\bf k},\sigma}$ creates a free (in vacuum) electron with momentum ${\bf k}$ and spin $\sigma$.
Also, $\varphi_{\bf k}=\tan^{-1}(k_y/k_x)+\pi/2$, $\cos\theta_{\bf k}=\Delta/E_k$ and $L^2$ is the area of the topological surface.

Let us consider a single electron-hole pair, where the electron is in the conduction band and the hole is in the valence band. 
Taking ${\bf P}=0$, the wavefunction for this electron-hole pair is
\begin{equation}
|{\rm xc}\rangle = \sum_{\bf k} \chi({\bf k}) \Psi^\dagger_{c,{\bf k}} \Psi_{v,{\bf k}}|0\rangle,
\end{equation}
where $|0\rangle$ is the ground state without particle-hole excitations, namely completely filled valence band and completely empty conduction band.
$\chi({\bf k})$ can be found from the requirement that $|{\rm xc}\rangle$ be an eigenstate of ${\cal H}$.
${\cal H}|{\rm xc}\rangle=\epsilon_{\rm ph}|{\rm xc}\rangle$ is satisfied if and only if 
\begin{equation}
\label{eq:sch}
(2 E_k+\delta\Sigma_k)\chi({\bf k})-\frac{1}{L^2}\sum_{\bf k'} V({\bf k},{\bf k}') \chi({\bf k'}) = \epsilon_{\rm ph} \chi({\bf k}),
\end{equation}
where $V({\bf k},{\bf k}')\equiv \langle c,{\bf k}; v,{\bf k'}|{\cal U}|c,{\bf k'}; v, {\bf k}\rangle=(2\pi e^2/\epsilon |{\bf k}-{\bf k'}|)\langle c,{\bf k}|c,{\bf k'}\rangle\langle v,{\bf k'}|v,{\bf k}\rangle$.
In the derivation of Eq.~(\ref{eq:sch}) we have treated the interband Coulomb interaction ($V^{(E)}$) exactly while adopting a mean-field Hartree-Fock approximation\cite{ando1997} for the intraband interaction ($V^{(A)}$). 
A more rigorous analysis requires solving the Bethe-Salpeter equation.\cite{rohlfing2000}
Moreover, we have normal-ordered the density operators ($:\Psi^\dagger_c \Psi_c:=\Psi^\dagger_c \Psi_c$ and $:\Psi^\dagger_v \Psi_v:=-\Psi_v \Psi^\dagger_v$), and have used $\Psi_c|0\rangle=0=\Psi^\dagger_v|0\rangle$.

The interband Coulomb matrix element in Eq.~(\ref{eq:sch}) reads
\begin{widetext}
\begin{equation}
\label{eq:coulomb}
V({\bf k},{\bf k}')= \frac{2\pi e^2}{2 \epsilon |{\bf k}-{\bf k'}|}\left[\frac{\hbar^2 v_F^2 k k'}{E_k E_{k'}}+\left(1+\frac{\Delta^2}{E_k E_{k'}}\right)\hat{\bf k}\cdot\hat{\bf k}'-i \Delta\left(\frac{1}{E_k}+\frac{1}{E_{k'}}\right) \hat{z}\cdot (\hat{\bf k}\times \hat{\bf k}')\right],
\end{equation} 
\end{widetext}
where $\epsilon$ is the momentum-dependent static dielectric constant at the surface of the TI.
A proper microscopic theory of $\epsilon$ requires the evaluation of a Lindhard function\cite{mahan} that includes contributions from both surface and bulk electronic bands. 
In lieu of a detailed knowledge of the full band structure, we hereafter regard $\epsilon$ as an empirical, momentum-independent quantity.
The assumption of momentum-independence will be justified {\rm a posteriori} (Figs.~(\ref{fig:wf1}) and ~(\ref{fig:wf2})), when we find that $\chi({\bf k})$ is peaked at $k\simeq 0$. 
Due to this, the most important Coulomb matrix elements in Eq.~(\ref{eq:sch}) carry a small momentum transfer ($|{\bf k}-{\bf k}'|\simeq 0$) and accordingly the long-wavelength limit of the static dielectric constant furnishes a satisfactory description of screening.
 
In Eq.~(\ref{eq:coulomb}), the second and last terms within the brackets are unconventional in the sense that they depend on the relative angle between the momenta of the interacting particles. 
In particular the last term is chiral, i.e. sensitive to the relative angular momentum of the interacting particles, and arises purely due to the broken time-reversal symmetry (it changes sign under $\Delta\to -\Delta$.)
For $\Delta=0$ the Coulomb matrix element agrees with that of graphene.
Note also that $V({\bf k},{\bf k}')=V^*({\bf k}',{\bf k})$.
Eq.~(\ref{eq:coulomb}) is the two-particle analog of the skew scattering that contributes to the anomalous Hall effect in ferromagnetic metals.\cite{nagaosa2010}
In that case the probability of scattering off a nonmagnetic impurity is dependent on the chirality or handedness between the initial and final momenta.
In our case, the strength with which an electron and a hole interact is skewed towards a particular sign of their relative angular momentum.
This has important implications for the exciton binding energies, as we discuss below.

Continuing with Eq.~(\ref{eq:sch}), the renormalization of the band-gap in the Hartree-Fock approximation\cite{ando1997} is given by the difference between self-energies of the conduction and valence band:
\begin{eqnarray}
\delta\Sigma_{\bf k}&=&-\frac{1}{L^2}\sum_{\bf k'} \frac{2\pi e^2}{\epsilon |{\bf k}-{\bf k'}|}\left(|\langle c,{\bf k}|v, {\bf k'}\rangle|^2-|\langle v,{\bf k}|v, {\bf k'}\rangle|^2\right)\nonumber\\
&=&\frac{1}{L^2}\sum_{\bf k'} \frac{2\pi e^2}{\epsilon |{\bf k}-{\bf k'}|}\frac{\Delta^2+\hbar^2 v_F^2 {\bf k}\cdot{\bf k'}}{E_k E_{k'}},
\end{eqnarray}
where a momentum cutoff $k_c$ must be introduced in order to regularize the ultraviolet divergence that reflects the breakdown of the low-energy effective theory.
$\delta\Sigma_{\bf k}$ does not depend on the direction of ${\bf k}$, and thus preserves the rotational symmetry of the problem.
$\delta\Sigma_{\bf k=0}=2 g \Delta \ln(2 k_c \hbar v_F/\Delta)$ renormalizes the gap at the Dirac point, and as expected it vanishes for $\Delta=0$.
For $k\gg\Delta/(\hbar v_F)$, $\delta\Sigma_{k}$ is generally nonzero even if $\Delta\to 0$.
In order to capture the physical role of the self energy, it is illustrative to represent it in operator language, 
\begin{eqnarray}
&&\delta\hat{\Sigma}=\sum_{\bf k}\frac{\delta\Sigma_{\bf k}}{2} (\Psi^\dagger_{c,{\bf k}}\Psi_{c,{\bf k}}-\Psi^\dagger_{v,{\bf k}} \Psi_{v,{\bf k}})\nonumber\\
&=&\sum_{\bf k}\frac{\delta\Sigma_{\bf k}}{2 E_k}\left(c^\dagger_{{\bf k},\uparrow},c^\dagger_{{\bf k},\downarrow}\right){\boldsymbol\sigma}\cdot(\Delta\hat{z}+v_F\hat{z}\times{\bf k})\left(\begin{array}{c} c_{{\bf k},\uparrow}\\c_{{\bf k},\downarrow}\end{array}\right)\nonumber,
\end{eqnarray}
which indicates that Coulomb interactions increase the bare Fermi velocity and the bare magnetic gap via
\begin{eqnarray}
\label{eq:ren}
v_F &\to& v_F (1+\delta\Sigma_k/2 E_k)\nonumber\\
\Delta &\to& \Delta(1+\delta\Sigma_k/2 E_k).
\end{eqnarray}
For $k\ll\Delta/(\hbar v_F)$ this enhancement amounts to up to 10\% of the bare value.
Eq.~(\ref{eq:ren}) suggests an explanation for why the Fermi velocity of the topological surface states in Bi$_2$Se$_3$ has been found to decrease\cite{wray2010} under Cu doping. 
In effect, the Coulomb interaction becomes weaker due to increased $\epsilon$ and onset of screening by doped carriers, and therefore the renormalized Fermi velocity is smaller than for the undoped case.
 
\subsection{Bound Electron-Hole Pairs}

Eq.~(\ref{eq:sch}) is a hydrogen-like Schrodinger equation for an electron-hole pair with zero COM momentum.
As such we expect a solution that will consist of a discrete set of bound states (excitons), as well as a continuum of scattering states.
Because of the unconventional momentum dependence of the Coulomb matrix elements in Eq.~(\ref{eq:coulomb}), Eq.~(\ref{eq:sch}) must be solved numerically.
We begin by decomposing the wavefunction into angular momentum channels $\chi({\bf k})=\chi_m(k)\exp(i m\phi)$, which leads to
\begin{equation}
\label{eq:2body}
(2 E_k+\delta\Sigma_k)\chi_m(k)-\int_0^\infty \frac{d k' k'}{2\pi} V_m (k,k')\chi_m(k') = \epsilon_{\rm ph} \chi_m(k),
\end{equation}
where 
\begin{equation}
\label{eq:vm}
V_m(k,k')=\int_0^{2\pi}\frac{d\phi}{2\pi} e^{i m\phi} V({\bf k},{\bf k}')
\end{equation}
 and we have used $(1/L^2)\sum_{\bf k}\to \int d{\bf k}/(2\pi)^2$. 
In Eq.~(\ref{eq:vm}) $\phi=\phi_{\bf k}-\phi_{{\bf k}'}$ is the angle between ${\bf k}$ and ${\bf k}'$.
While Eq.~(\ref{eq:vm}) is physically meaningful only for $k,k'<k_c$, the rapid decay of the electron-hole wavefunction in momentum space (see below) justifies taking the entire range of momenta.
It is clear from Eq.~(\ref{eq:2body}) that $m$ is a good quantum number for electron-hole pairs with vanishing COM momentum.
This is the U(1) symmetry identified in the previous section, and accordingly $m$ is the projection of the total angular momentum along the $z$-axis . 
We solve Eq.~(\ref{eq:2body}) numerically using a modified quadrature method\cite{chao1991} that carefully treats the singularity of the Coulomb matrix element at ${\bf k}={\bf k}'$ (see Appendix).

\begin{figure}[h]
\begin{center}
\includegraphics[scale=0.3,angle=270]{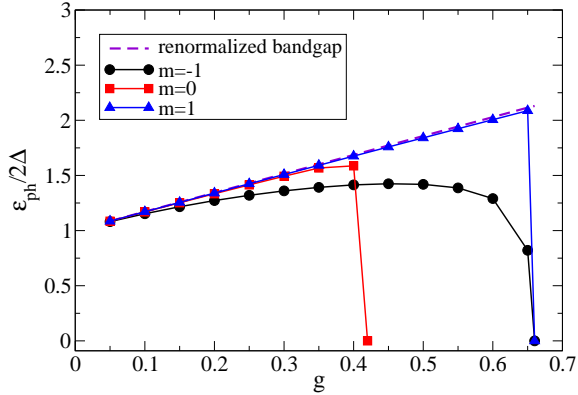}
\caption{Excitation energies as a function of the dimensionless Coulomb coupling $g=(e^2/\epsilon)/(2\hbar v_F)$, for an electron-hole pair located at the surface of a TI with a magnetic gap.  
 $m$ is the quantum number for the projection of the total angular momentum along the $z$-axis and $E_g=2\Delta+\delta\Sigma_{{\bf k}=0}$ is the renormalized energy gap. As derived in the text, the renormalized gap grows linearly with $g$, and its precise magnitude depends logarithmically on an ultraviolet momentum cutoff. The exciton binding energy is given by $E_g-\epsilon_{\rm ph}$ and is found to obey Eq.~(\ref{eq:topo}).  For each $m$, only the lowest-lying bound state is shown. In the weak coupling regime (which is relevant for most TIs discovered to date) the $m=-1$ exciton has the largest binding energy. However, the $m=0$ channel is most prone to an excitonic instability (at $g\simeq 0.4$). The critical value of $g$ for the instability can be roughly estimated from Eq.~(\ref{eq:topo}) via $\epsilon_b (g\to g_c)\simeq 2\Delta$, or else from dimensional arguments outlined in the text.}
\label{fig:eb}
\end{center}
\end{figure}

\begin{figure}[h]
\begin{center}
\includegraphics[scale=0.3,angle=270]{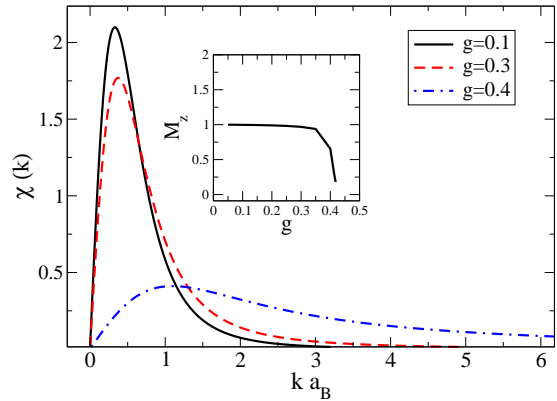}
\caption{Momentum space wave function for the lowest-lying electron-hole pair with $m=0$.
$a_B=\hbar v_F/(\Delta g)$ is an effective Bohr radius (the effective mass in the parabollic band approximation is $\Delta/2 v_F^2$).  The wavefunction (normalized through $\int_{0}^\infty d\kappa \kappa |\chi_m(\kappa)|^2=1$, where $\kappa=k a_B$)  is peaked at $k a_B\sim O(1)$ ($\hbar v_F k/ \Delta\sim g\ll1$), and its spread in momentum space increases as interactions become stronger. {\em Inset:} the z-component of the spin density for the same exciton state. Weakly interacting particle-hole pairs are spin polarized along the direction normal to the surface. This spin projection decreases abruptly in the vicinity of the excitonic instability.}
\label{fig:wf1}
\end{center}
\end{figure}

\begin{figure}[h]
\begin{center}
\includegraphics[scale=0.3,angle=270]{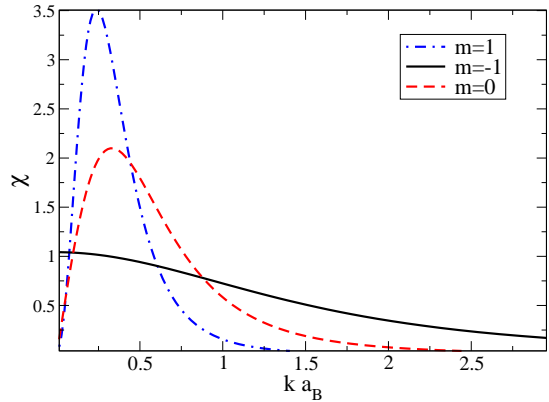}
\caption{Momentum space wave function for the lowest-lying electron-hole pair with $g=0.1$ and $m=1,0,-1$. The $m=-1$ exciton wavefunction is more spread out in momentum space, which is partly why the corresponding binding energy is larger.}
\label{fig:wf2}
\end{center}
\end{figure}

In Fig.~(\ref{fig:eb}) we plot the energy of the lowest lying exciton states for different values of $m$.
At weak Coulomb coupling the binding energy is largest for $m=-1$, and the excitation energies for $m=1$ and $m=-1$ are different.
Such asymmetry in the sign of the angular momentum stems from the aforementioned chirality of the Coulomb matrix element (Eq.~(\ref{eq:coulomb})), which enhances the electron-hole attraction for angular momenta aligned with the Zeeman field $\Delta\hat{z}$. 
As such, $m$ and $-m$ are interchangeable via $\Delta\to -\Delta$.
This is all in stark contrast to conventional hydrogen-like excitons.
In ordinary two-dimensional excitons with a reduced effective mass $m_{\rm eff}$, the binding energy is given by $\epsilon_b=E_0/(n+1/2)^2$, where $n=0,1,2,...$, $E_0=(m_{\rm eff}/m_e \epsilon^2) 13.6 {\rm eV}$ and $m_e$ is the bare electron mass.
It is instructive to write down the expression for the largest binding energy of conventional excitons with a given value of the angular momentum $m$: 
\begin{equation}
\label{eq:conv}
\epsilon_{b,m}^{\rm max}=\frac{E_0}{(|m|+1/2)^2} \mbox{   (ordinary 2D excitons)}.
\end{equation}
The decrease in the binding energy of the strongest bound exciton with $m$ may be interpreted by invoking a centrifugal potential that reduces the effective attraction between the electron and the hole for higher angular momentum channels.
In contrast to Eq.~(\ref{eq:conv}), our numerical analysis for excitons on topological surfaces gives
\begin{equation}
\label{eq:topo}
\epsilon_{b,m}^{\rm max}=\frac{\tilde{E}_0}{(|m+1|+1/2)^2} \mbox{   (topological 2D excitons)},
\end{equation}
where $\tilde{E}_0=\alpha\Delta g^2$ and $\alpha$ is a constant of order one. 
$\Delta g^2$ is the Coulomb energy scale corresponding to electron-hole pairs with a reduced effective mass $\Delta/2 v_F^2$.
This mass agrees with the inverse of the non-interacting band curvature at ${\bf k}=0$.
The remarkable difference between Eq.~(\ref{eq:conv}) and Eq.~(\ref{eq:topo}) is a shift in the angular momentum quantum number (i.e. $m\to m+1$), which is consistent with the nontrivial Berry phase of the topological surface states.
Given their maximal binding energy for $m=-1$ as well as their asymmetry under $m\to -m$, the excitons occurring at topological surfaces with magnetic order can be described as  ``p-wave'' and ``chiral''.
As we show below, these unconventional features may be optically detectable.
 
For $g\simeq 0.1$ and $\Delta\simeq 20 {\rm meV}$, the binding energy for the lowest-lying exciton is about $1 {\rm meV}$.
Though small, this binding energy is comparable to that of biexcitons in GaAs quantum wells, which have been seen in photoluminescence experiments.\cite{miller1982} 
In the next subsection we show that excitonic features in the optical absorption spectrum can be noticeable even for $g\simeq 0.05$ and $\Delta\simeq 10{\rm meV}$.

For larger Coulomb coupling ($g\gtrsim 0.4$), the binding energy for the $m=0$ exciton surpasses that of the $m=-1$ exciton, and leads to an excitonic instability towards an ``s-wave'' Bose condensate.
The critical value of $g$ for the onset of the excitonic instability can be estimated from dimensional analysis.
At zero temperature and when the Fermi energy is in the gap, the only lengthscale\cite{low energy} for non-interacting electrons is $\xi_\Delta=\hbar v_F/\Delta$, whereas the only lengthscale related to Coulomb interactions is $\xi_e=e^2/(\epsilon\Delta)$.
The validity of our calculations is limited to the weakly interacting regime where $\xi_e\ll\xi_\Delta$.
In contrast, when $\xi_e\gg\xi_\Delta$ the single-exciton approximation breaks down and the physics is governed by many-body interactions.
Therefore the excitonic instability occurs at $\xi_e\simeq\xi_\Delta$, which yields $g_c\simeq 0.5$.
This value agrees well with the numerical results of Fig.~(\ref{fig:eb}). 
For $v_F\simeq 4\times 10^{5} {\rm m/s}$, $g=2.72/\epsilon$ and thus the excitonic instability is reached at $\epsilon\simeq 7$. 
At present we are not aware of any TI that has such a small value of the dielectric constant at its surface.

Fig.~(\ref{fig:wf1}) illustrates the wavefunction for the $m=0$ exciton in momentum space.
Most of its weight is concentrated at $k a_B\lesssim O(1)$, where $a_B=\hbar v_F/(\Delta g)$ is an effective Bohr radius.
For strongly interacting electron-hole pairs, the wavefunctions become more delocalized in momentum space.
This trend reflects the tendency of the electron and the hole to stay close to one another in real space, so as to optimize the Coulomb interaction. 
The inset of Fig.~(\ref{fig:wf1}) shows the spin density of the $m=0$ exciton.
In the non-interacting case, the spinor of the conduction band at momentum {\bf k} is perfectly antialigned with respect to that of the valence band. 
Therefore, non-interacting particle-hole excitations are spinful.
The $x$- and $y$-components of the spin density vanish by symmetry, while the $z$-component of the spin density for electron-hole pair with angular momentum $m$ is
\begin{equation}
\label{eq:Mz}
M_z = \frac{1}{2}\langle{\rm xc}|\sigma_e^z+\sigma_h^z|{\rm xc}\rangle
= \sum_n\int\frac{dk k}{2\pi}\frac{\Delta}{E_k}|\chi_{nm}(k)|^2,
\end{equation}
where $n$ labels the different eigenstates for a given value of $m$.
As evidenced by Fig.~(\ref{fig:wf1}), $M_z$ varies little as a function of $g$ up until $g\simeq g_c$.
It is sensible that the spin-polarization of electron-hole pairs be largely immune to weak Coulomb interactions, because the latter are spin-rotationally invariant. 

In Fig.~(\ref{fig:wf2}) we plot the momentum space wave functions for the $m=1,0,-1$ excitons. 
The $m=-1$ ($m=1$) eigenstate is most (least) delocalized in momentum space.
This is a consequence of the chirality in the Coulomb interaction, which makes the electron-hole attraction largest for $m=-1$ and smallest for $m=1$.
Hence the $m=-1$ wavefunction is more delocalized in momentum space.

\subsection{Optical Absorption}

The knowledge of particle-hole excitation energies and wavefunctions at $P=0$ provides a gateway for determining the optical conductivity\cite{yu2001,haug2005} of topological surface states whose chemical potential is placed inside a magnetic energy gap.
From Fermi's golden rule, the optical absorption for photons of frequency $\omega$ is
\begin{equation}
\label{eq:r}
R(\omega)=\frac{2\pi}{\hbar}\sum_{\alpha}|\langle \alpha|\frac{e}{c}{\bf A}\cdot{\bf v}|0\rangle|^2 \delta(\epsilon_{\rm ph}^{(\alpha)}-\hbar\omega),
\end{equation}
where $\alpha$ is the set of quantum numbers describing a particle-hole excitation and $|0\rangle$ is the ground state in absence of particle-hole excitations.
Ignoring electric fields (which are included in the next section) we have $\alpha\to(n, m)$, where $m$ refers to the angular momentum quantum number and $n$ counts the different eigenstates for a given $m$.
${\bf v}=\partial{\cal H}_0/\hbar\partial{{\bf k}}=v_F ({\boldsymbol\sigma}\times\hat{z})$  is the velocity operator for non-interacting electrons (we neglect the renormalization of the Fermi velocity due to interactions), ${\bf A}$ is the vector potential and
\begin{equation}
\label{eq:dipolo}
\langle \alpha|v^i|0\rangle =\sum_{\bf k} \chi_{n m}(k) e^{i m\phi} \langle c {\bf k}|v^i|v {\bf k}\rangle
\end{equation}
is the electric dipole transition matrix element.

\begin{figure}[h]
\begin{center}
\includegraphics[scale=0.3,angle=270]{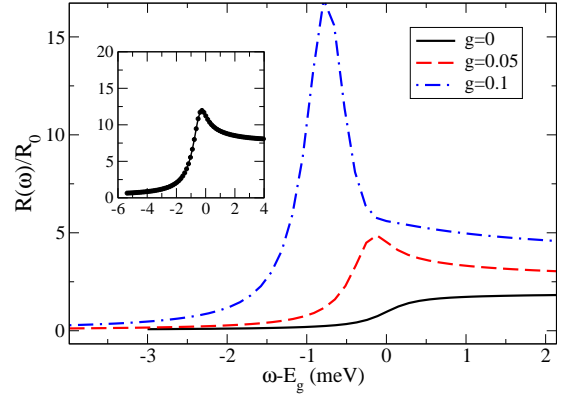}
\caption{Optical absorption for linearly polarized light, as a function of the dimensionless Coulomb coupling $g$. $\Delta=20{\rm meV}$, $\gamma=0.6 {\rm meV}$, and $R_0$ is the absorption coefficient for free electron-hole pairs at $\omega=2\Delta$. $E_g=2\Delta+\delta\Sigma_{{\bf k}=0}$ is the renormalized energy gap.  Most of the subgap absorption originates from $m=-1$ (or $m=1$ if one reverses the direction of magnetization). {\em Inset:} Optical absorption spectrum for linearly polarized light, with $g=0.1$, $\Delta=10 {\rm meV}$ and $\gamma=1.2 {\rm meV}$. The intensity of the $m=-1$ intensity peak is reduced but still noticeable.}
\label{fig:oa1}
\end{center}
\end{figure}

\begin{figure}[h]
\begin{center}
\includegraphics[scale=0.3,angle=270]{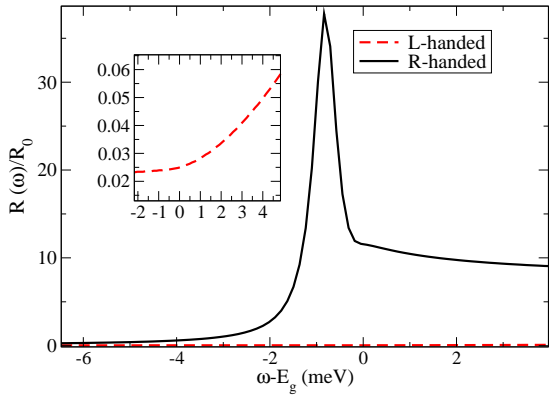}
\caption{Optical absorption for circularly polarized light. $g=0.12$, $\Delta=15{\rm meV}$ and $\gamma=0.6{\rm meV}$. $R_0$ is the absorption coefficient for $g=0$ and $\omega=2\Delta$, taken as the average of left-circularly and right-circularly polarized lights.  $E_g=2\Delta+\delta\Sigma_{{\bf k}=0}$ is the renormalized energy gap. For right-circularly polarized light the most strongly bound exciton ($m=-1$) is optically bright. For left-circularly polarized light $m=-1$ is optically dark and the absorption spectrum is missing a prominent excitonic peak. The right- and left-circularly polarized spectra map onto one another under $\Delta\to -\Delta$, namely under switching of the magnetization. {\em Inset:} Zoomed in absorption spectrum for the left-circularly polarized light. The coherence factor $\sin^2(\theta/2)$, mentioned in the text, diminishes the absorption intensity considerably.}
\label{fig:oa2}
\end{center}
\end{figure}

Because $[{\bf v},{\cal H}_0]\neq 0$, the velocity operator induces interband transitions and is able to produce magnetized excitons.
The optical absorption vanishes for ${\bf A}=A\hat{z}$ and is symmetric in the $xy$ plane.
For linearly polarized light (e.g. ${\bf A}=A\hat{x}$, though one can immediately verify that any linear polarization in the $xy$ plane yields an identical optical absorption), the dipole matrix element is
\begin{equation}
\label{eq:coh}
\langle c {\bf k}| v^x| v {\bf k}\rangle = -i v_F\left(e^{i\varphi_{\bf k}}\cos^2\frac{\theta_{\bf k}}{2}+e^{-i\varphi_{\bf k}}\sin^2\frac{\theta_{\bf k}}{2}\right),
\end{equation}
where $\theta_{\bf k}=\cos^{-1}(\Delta/E_{\bf k})$ and $\varphi_{\bf k}=\phi + \pi/2$ are the spinor angles introduced in Section III.
It follows from Eqs.~(\ref{eq:dipolo}) and (\ref{eq:coh}) that only $m=1$ and $m=-1$ states are optically active for linearly polarized light, whereas the $m=0$ exciton is dark.
This selection rule differs from ordinary hydrogen-like excitons and even from excitons in many topologically trivial spin-orbit coupled systems, wherein $m=0$ is a bright exciton.
It is likewise different from the optical selection rules in gated bilayer graphene.\cite{park2010}
For circularly polarized light (${\bf A}=A(\hat{x}\pm i\hat{y})/\sqrt{2}$) we have
\begin{eqnarray}
\label{eq:coh2}
\langle c {\bf k}| v^+| v {\bf k}\rangle &=& -\sqrt{2} i v_F e^{i\varphi_{\bf k}}\cos^2\frac{\theta_{\bf k}}{2}\nonumber\\
\langle c {\bf k}| v^-| v {\bf k}\rangle &=& -\sqrt{2} i v_F e^{-i\varphi_{\bf k}}\sin^2\frac{\theta_{\bf k}}{2},
\end{eqnarray}
where $v^\pm=(v^x\pm i v^y)/\sqrt{2}$.
Thus only the $m=-1$ ($m=1$) exciton is optically active for right- (left-) circularly polarized light.
In Figs.~(\ref{fig:oa1}) and ~(\ref{fig:oa2}) we plot the optical absorption for linearly and circularly polarized light.
We incorporate the finite lifetime of the excitons by broadening the Dirac delta function $\delta(\epsilon_{n,m}-\hbar\omega)$ into a Lorentzian 
\begin{equation}
\delta(\epsilon_{\rm ph}^{(n,m)}-\hbar\omega)\rightarrow\frac{\gamma/2}{(\epsilon_{\rm ph}^{(n,m)}-\hbar\omega)^2+\gamma^2/4},
\end{equation}
where $\gamma$ is the linewidth.
We have taken $\gamma\simeq 0.6 {\rm meV}$, which is fairly small but commonly obtained in a variety of semiconducting quantum wells.\cite{kim1992}

In the non-interacting case the optical absorption increases linearly at frequencies above the gap; this is simply a manifestation of the Dirac-like density-of-states.
For frequencies in the vicinity of the gap, electron-hole interactions enhance the optical absorption.
When the surface is irradiated with either linearly polarized or right-circularly polarized light, an excitonic absorption peak is visible even in samples with modest gaps ($\Delta\simeq 10 {\rm meV}$) and weak interactions ($g\simeq 0.05$ or $\epsilon\simeq 50$).
Part of the reason for this prominent excitonic peak is that the ``coherence factor'' $\cos^2(\theta/2)$ in Eqs.~(\ref{eq:coh}) and ~(\ref{eq:coh2}) has most of its weight at $\hbar v_F k\ll\Delta$, which is precisely the region of momentum space where the exciton wavefunction is peaked (recall Fig.~(\ref{fig:wf1})). 

In case of left-circularly polarized light, the optically bright $m=1$ excitons are esentially merged with the continuum at experimentally relevant Coulomb interaction strengths (recall Fig.~(\ref{fig:eb})), and moreover the coherence factor $\sin^2(\theta/2)$ in the second line of Eq.~(\ref{eq:coh2}) is largely depleted in the region of momentum space where the excitonic wavefunction is localized. 
Consequently, the absorption spectrum of left-circularly polarized light shows no significant excitonic peaks and the absolute intensity of absorption near the bandgap is much smaller (Fig.~(\ref{fig:oa2})). 
Such a large asymmetry between left- and right-circularly polarized light should be experimentally testable.
Keeping in mind that right- and left-circularly polarized absorption spectra map onto one another under $\Delta\to -\Delta$, one simple way to detect the asymmetry between them is by shining circularly polarized light and then switching the direction of magnetization at the surface of the topological insulator by 180$^\circ$.

We finalize this section with a few caveats on how to facilitate the experimental detection of the aforementioned phenomena.
On one hand, the magnetic overlay must be a charge insulator with an optical gap that is large compared to the energy gap of the surface states. 
This requirement is clearly fulfilled in magnetic ions (e.g. Fe$^{3+}$, Mn$^{2+}$) or in magnetic semiconductors such as\cite{santos2008} EuO, which have $O(1{\rm eV})$ bandgaps. 
On the other hand, light can excite spatially uniform magnons in the ferromagnetic overlay.
However, the typical energy scale for these collective excitations\cite{vonsovski1966} is $0.05{\rm meV}\ll2\Delta$, and therefore the magnon absorption lines will be well separated from excitonic features.
Perhaps the most convenient system for testing our predictions is Bi$_2$Se$_3$ with Fe and Mn ions doped in the bulk or at the surface.\cite{chen2010} Bulk Mn doping leads to $\Delta\sim 7$meV at 1\% doping; substantially larger magnetic gaps have been reported through increase in the dopant concentration. 





\section{Electron-hole pairs on a one-dimensional surface}

Thus far we have considered two dimensional surfaces with translational and rotational symmetry.
In this section we concentrate on a one-dimensional variant of the interacting two-body problem (Eq.~(\ref{eq:h1})), which is pertinent to the surface of a quasi-one-dimensional TI.
The motivation for this investigation lies in the well-known fact that excitonic effects are stronger in reduced dimensions.
Besides, the one-dimensional model is numerically simple enough to allow us to learn with relative ease how excitonic effects are altered under external electric fields as well as in presence of a finite COM momentum for the electron-hole pairs.
In regards to the latter, some of the results derived in this section (e.g. Eq.~(\ref{eq:vg})) are relevant for two dimensional surfaces.

The low-energy band structure for the surface states of a cylindrical TI nanowire\cite{rosenberg2010} is
\begin{equation}
\label{eq:cyl}
E_{k,m}=\pm\hbar v_F \sqrt{k^2+\frac{(m+1/2-\nu)^2}{r_0^2}},
\end{equation}
where $k$ is the momentum along the axis of the cylinder, $m$ is the azimuthal projection of the total angular momentum, $\nu$ is the magnetic flux (in units of the flux quantum) that threads the cross section of the nanowire, and $r_0$ is the radius of the wire.
For $\nu=0$, the energy spectrum at the surface is gapped and all energy levels are doubly degenerate ($E_{k,m=-1}= E_{k,m=0}$).
For $\nu=1/2$, $m=0$ constitutes a gapless, non-degenerate, topological surface state.
When $\nu\in(0,1/2)$ the surface spectrum is gapped but the lowest-lying energy states are non-degenerate; this is the situation for which the calculation of the present section applies.
It follows from Eq.~(\ref{eq:cyl}) that the bandgap for non-interacting electrons is $2 \Delta=(2 \hbar v_F/r_0)|\nu-1/2|$, which is tunable by an external magnetic field along the axis of the cylinder.
Another way to open a uniform energy gap is to coat the nanowire with a radially magnetized ferromagnet, though this is experimentally cumbersome.
$\nu\simeq 1/2$ is achievable with magnetic fields $B\lesssim 10{\rm T}$ even for radii as small as $r_0\simeq 10 {\rm nm}$.
When $\nu\simeq 1/2$, it is a good approximation to focus on the $m=0$ state only, because it is well separated (at least near $k=0$) from higher energy states.   
For instance,  $E_{k=0,m=-1}=3 E_{k=0,m=0}$ for $\nu=1/4$.
For $\nu\simeq 1/4$ and $r_0\simeq 10 {\rm nm}$, $\Delta\sim O(10{\rm meV})$ is comparable to the magnitude of the exchange field considered in the previous section.

When ${\cal E}\neq 0$ the problem at hand contains no symmetries other than the conservation of the COM canonical momentum $P$.
 Hence it is more convenient to solve the two-body Dirac equation in real space.
Our starting model for the surface of a TI nanowire with radius $r_0$ and length $L$ is
\begin{eqnarray}
\label{eq:h1d}
{\cal H} &=& -v_F\left(\sigma_e^x-\sigma_h^x\right)P +\frac{v_F}{2}\left({\boldsymbol\sigma}_e+{\boldsymbol\sigma}_h\right)\cdot\left(-p\hat{x}+\frac{\Delta}{v_F}\hat{\Omega}\right)\nonumber\\
&-&\frac{e^2}{\epsilon (|y| + \eta r_0)}-e {\cal E} y,
\end{eqnarray}
where we have defined the axis of the cylinder to be the y-axis and $y\in(-L,L)$ is the relative coordinate along the wire.
The regularized form of the Coulomb interaction in Eq.~(\ref{eq:h1d}) can be derived by averaging the three-dimensional Coulomb potential with radial envelope functions.\cite{haug2005}
$\eta$ is a positive fitting parameter that takes the value of $0.3$ in certain idealizations of wires.

We solve Eq.~(\ref{eq:h1d}) as an eigenvalue problem in real space, with the boundary conditions chosen so that the wavefunction vanishes at the ends of the wire.
Starting from the non-interacting problem with $P={\cal E}=e^2/\epsilon=0$,  we keep only a quarter of eigenvalues and eigenstates.
These correspond to particle-hole pairs with positive excitation energies $\epsilon_{\rm ph}\geq 2\Delta$, namely a hole in the valence band and an electron in the conduction band.
Thereafter we analyze how these energy levels (and eigenstates) change under Coulomb interactions, electric fields and COM momenta.
The eigenvalues and eigenstates we discard describe (i) a valence band hole and a valence band electron, (ii) a conduction band electron and a conduction band hole, (iii) a conduction band hole and a valence band electron.

\begin{figure}[h]
\begin{center}
\includegraphics[scale=0.3,angle=270]{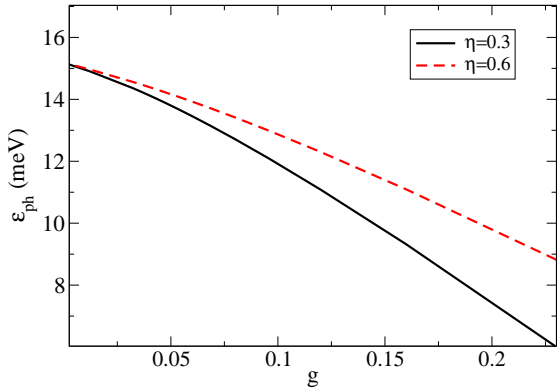}
\caption{Lowest-lying particle-hole excitation energy for a TI nanowire of radius $r_0=10 {\rm nm}$, length $L=100 {\rm nm}$ and $\hbar v_F=300 {\rm meV} {\rm nm}$. In our numerical calculation, each discrete step along the wire equals $1$ nm. The magnetic flux along the axis of the wire is $\nu=1/4$ in units of the flux quantum, which leads to a non-interacting bandgap of $2\Delta=15 {\rm meV}$.
For smaller values of the fitting parameter $\eta$, the exciton binding energies become larger.}
\label{fig:1d1}
\end{center}
\end{figure}

In Fig.~(\ref{fig:1d1}) we plot the lowest lying particle-hole excitation energy as a function of $g$, for ${\cal E}=\Omega_x=\Omega_y=P=0$.
Narrower wires show larger exciton binding energies: for wires of radius $\sim 10 {\rm nm}$ they may be $\sim 3-5$ times larger than the binding energies of the two dimensional excitons studied in the previous section.

\begin{figure}[h]
\begin{center}
\includegraphics[scale=0.3,angle=270]{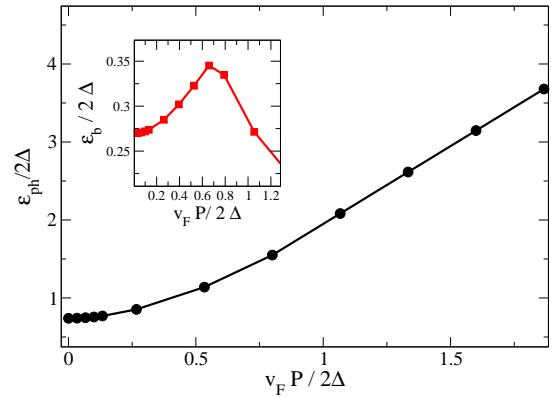}
\caption{Lowest-lying particle-hole excitation energy as a function of the canonical COM momentum $P$. $L=100 {\rm nm}$, $\Delta_-=0$ and $g=0.12$.
The dependence of the particle-hole energy on $P$ is quadratic at $P\ll\Delta/v_F$ (light excitons) and becomes linear as $P\gg\Delta/v_F$ (heavy excitons). 
When $\Delta_-\neq 0$, the dispersion is shifted.
A nearly identical dispersion is obtained (not shown) as a function of a radial exchange field $h_y$ that couples to $\sigma_e^y+\sigma_h^y$. The effect of this exchange field is to renormalize the bandap via $\Delta\to \sqrt{\Delta^2+h_y^2}$. Azimuthal exchange fields (which couple to $\sigma_e^x+\sigma_h^x$) amount to a redefinition of the origin in momentum space and do not change the particle-hole excitation spectrum. {\rm Inset:} The exciton binding energy has a non-trivial dependence on $P$.}
\label{fig:1d2}
\end{center}
\end{figure}

In Fig.~(\ref{fig:1d2}) we display the influence of $P\neq 0$ and radial exchange fields in the particle-hole excitation energy (azimuthal exchange fields amount to a shift in $p$ and are inconsequential).
The binding energy has a non-monotonic dependence on $P$; this is markedly different from conventional systems wherein the COM and relative coordinates are decoupled and thus the exciton binding energy is independent of $P$ (at least in absence of quantizing magnetic fields).
For $P\ll\Delta/v_F$, excitons are formed predominantly out of electrons (holes) located near the bottom (top) of the non-interacting conduction (valence) band. 
Accordingly we find $\epsilon_{\rm ph}(P)\simeq\epsilon_{\rm ph}(0)+P^2/(2 M)$, where the exciton mass is given by $M=\Delta/(2 v_F^2)$.
As $P$ increases, the particles and holes forming the exciton necessarily explore the linear part of the non-interacting energy-momentum dispersion. 
Thereby excitons become heavier; in fact for  $P\gg\Delta/v_F$ we obtain $\partial\epsilon_{\rm ph}/\partial P\simeq 2 v_F$, which translates to infinitely massive particle-hole pairs.


\begin{figure}[h]
\begin{center}
\includegraphics[scale=0.3,angle=270]{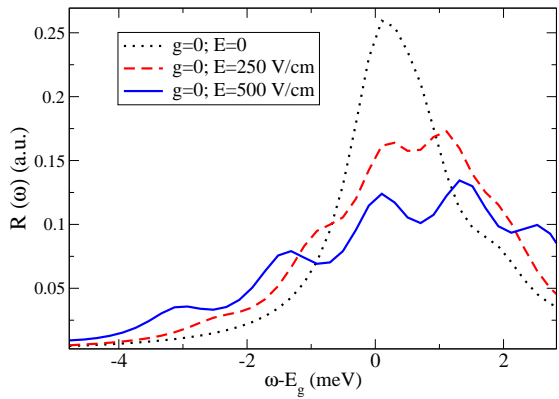}
\caption{Optical absorption (in arbitrary units) for non-interacting electrons illuminated by linearly polarized light (with its polarization vector along the axis of the nanowire). $L=150 {\rm nm}$ is the length of the nanowire, $2 \Delta=15 {\rm meV}$ is the non-interacting bandgap, $\nu=1/4$ is the magnetic flux threading its cross section, $\eta=0.3$ is the fitting parameter that enters in the regularized Coulomb interaction,  and $\gamma=1.2 {\rm meV}$ is the excitonic linewidth. In absence of electric fields, the absorption spectrum reflects the one-dimensional density of states (modulated by the ``coherence factors'' of Eq.~(\ref{eq:dip})) at frequencies above the bandgap. In presence of electric fields the subgap absorption is enhanced and Franz-Keldysh oscillations appear.} 
\label{fig:1d3}
\end{center}
\end{figure}

\begin{figure}[h]
\begin{center}
\includegraphics[scale=0.3,angle=270]{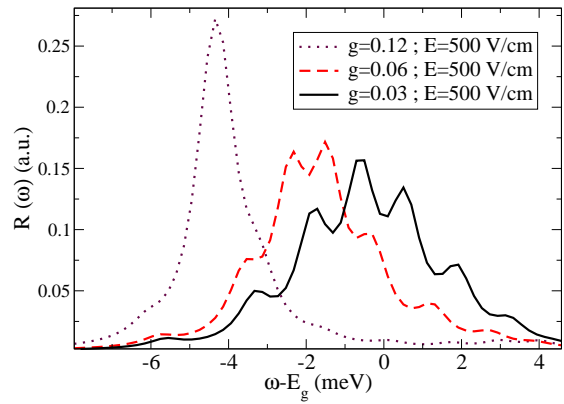}
\caption{Optical absorption for interacting electrons illuminated by linearly polarized light. The parameters of the nanowire are identical to those of the previous figure, with $\gamma=1$ meV. $E_g=2\Delta$ is the bandgap for non-interacting electrons. When ${\cal E}=500 {\rm V/cm}$, the influence of the electric field in the optical absorption is clearly visible for $g=0.03$ and less discernible for $g=0.12$.}
\label{fig:1d4}
\end{center}
\end{figure}

The observations from the preceding paragraph are also applicable to the two-dimensional magneto-excitons discussed in Section II.
Yet unlike in nanowires, in two-dimensional topological surfaces the magnetic gap is generated by exchange interactions.
These exchange interactions are generally asymmetric between electrons and holes. 
An interesting consequence of this asymmetry is that it introduces a mechanical COM momentum ${\boldsymbol\Pi}={\bf P}-\Delta_-/v_F\hat{z}\times\Omega$, where $\Delta_-$ denotes the difference in the exchange fields between electrons and holes (see Section II for the definition of $\Delta_\pm$).
When $\Delta_-\neq 0$, Fig.~(\ref{fig:1d2}) remains valid insofar as $P$ is replaced by $\Pi$ in the horizontal axis.
As a result optically generated excitons (i.e. $P=0$ excitons) carry a nonzero COM group velocity in presence of in-plane exchange fields:
\begin{equation}
\label{eq:vg}
 {\bf v}_g=\lim_{P\to 0}\frac{\partial\epsilon_{\rm ph}({\bf P})}{\partial {\bf P}}= v_F \frac{\Delta_-}{\Omega_z\Delta_+}\hat{z}\times\hat\Omega. 
\end{equation}
Eq.~(\ref{eq:vg}) was derived for $\Delta_-|\hat{z}\times\hat\Omega|\ll\Omega_z\Delta_+$, and yields $v_g<<v_F$. 
 This nonzero group velocity of optically excited excitons, combined with their nonzero spin polarization along the z-axis (see Eq.~(\ref{eq:Mz}) and the discussion therein), opens prospects for creating excitonic spin currents and spin Hall effects.
Similarly, by tilting $\hat\Omega$ away from $\hat{z}$ one can map the excitonic dispersion relation (Fig.~(\ref{fig:1d2})) optically. \cite{lozovic2002}  

Finally, we investigate the photoluminescence  (Eq.~(\ref{eq:r}))  of the nanowire in presence of electric fields.
Let us denote the eigenstates of Eq.~(\ref{eq:h1d}) as $\chi_\alpha^{\sigma\sigma'} (y)$, where the index $\alpha$ is such that $\epsilon_{{\rm ph}}^{(\alpha)}\geq 2\Delta$ in the non-interacting case (recall the discussion above), and $\sigma$ ($\sigma'$) is the spin label of the electron (hole).
Then the dipole matrix element is given by
\begin{equation}
\label{eq:dip}
\langle\alpha|v^y|0\rangle=\sum_k\sum_{\sigma\sigma'\in\{\uparrow,\downarrow\}}\chi_\alpha^{\sigma\sigma'}(k) {\cal M}_{\sigma\sigma'}(k)\langle c k|v^y | v k\rangle,
\end{equation}
where $v^y$ is the velocity operator along the wire, $\chi_\alpha (k)=\int_{-L}^L d y e^{- i k y} \chi_\alpha (y)$ and
 ${\cal M}_{\sigma\sigma'}=\Psi_{e\sigma}(k)\otimes\Psi_{h\sigma'}(-k)=(\cos(\theta_k/2),-\sin(\theta_k/2))\otimes (\cos(\theta_k/2),-\sin(\theta_k/2))$.
The optical absorption is nonzero provided that the polarization of the impinging light has a nonzero projection along the axis of the nanowire.
Fig.~(\ref{fig:1d3}) shows that electric fields enhance the subgap optical absorption due to particle-hole excitations with $\epsilon_{\rm ph}<2\Delta$ being formed between different points along the wire.
The electric-field-induced oscillations in photoluminescence, i.e. Franz-Keldysh oscillations,\cite{yu2001,haug2005} are visible for electric fields ${\cal E}\gtrsim 250 {\rm V/cm}$.
Fig.~(\ref{fig:1d3}) illustrates that the influence of a given electric field in the excitonic absorption peak decreases as the Coulomb interactions are made stronger.
 This is reasonable because the electron and the hole tend to stay close to one another in order to maximize their Coulomb attraction, whereas the electric-field term in the Hamiltonian vanishes when the electron and the hole are located at the same point in space. 
Incidentally, we find that an electric field induces no net spin polarization of the exciton.

\section{Conclusions}

Motivated in part by the objective to understand the interplay between Coulomb and spin-orbit interactions on topological insulators with broken time reversal symmetry, 
we have studied the excitonic photoluminescence on a magnetized surface of a strong TI. 
Throughout this study we have encountered a number of potentially interesting properties.
First, optically excited electron-hole pairs are magnetized in the direction perpendicular to the topological surface.
Second, the relative motion of an electron-hole pair is entangled to the center-of-mass motion.
Accordingly, the excitonic Berry phase is nonzero and a spin-Hall effect of excitons ensues.
Third, Coulomb interactions enhance both the Fermi velocity and the mass of Dirac fermions. 
Fourth, in two dimensional topological surfaces, the combination of spin-orbit interaction and broken time reversal symmetry leads to a skewness or chirality in the effective Coulomb interaction between an electron and a hole.
The underlying cause for this chirality resides in the nonzero Berry phase of the topological surface states.
Consequently, the strongest bound state occurs for an electron-hole pair whose total angular momentum has a projection $\hbar$ (or $-\hbar$, depending on the direction of magnetization at the surface) along the normal to the surface.
This ``p-wave'' and ``chiral'' exciton becomes optically bright upon shining circularly polarized light with the proper handedness, and is optically dark upon shining of circularly polarized light with the opposite handedness. 
For fixed handedness of the impinging light, bright-to-dark transitions can be induced by switching the direction of magnetization by $180^\circ$.
Fifth, so long as there is an electron-hole asymmetry in the exchange coupling between topological surface states and magnetic ions,  optically excited electron-hole pairs have a nonzero center-of-mass group velocity that may be tuned by tilting the magnetization away from the normal of the surface.  

Many of the anticipated peculiarities may be experimentally accessible in topological surfaces that host weak-to-moderate Coulomb interactions ($g=(e^2/\epsilon)/(2\hbar v_F)\gtrsim 0.1$), promote sufficiently long exciton lifetimes, and have their chemical potential in the energy gap (both in bulk and at the surface). 
If the chemical potential intersects the topological surface bands, an optical gap will certainly be still present. 
Nonetheless, excitonic effects will be substantially weakened because exciton binding energies\cite{ion} and lifetimes\cite{gavoret} both decrease rapidly in the metallic regime.  

Aside from the task of characterizing magneto-excitons that form between Landau levels under perpendicular magnetic fields, we can envision various directions along which the work presented here can be expanded. 
On one hand, it will be interesting to consider ultrathin TI ribbons, wherein two Dirac cones (one on each surface) and spatially indirect excitons might contribute to photoluminescence.
On the other hand, full-fledged electronic structure calculations of the topological surface states will be crucial\cite{caveat} in order to make quantitative contact with optical measurements that are likely to be performed in the foreseeable future.

\acknowledgements
I.G. acknowledges useful comments from I. Affleck. This research has been financially supported by a Junior Fellowship of the Canadian Institute for Advanced Research (I.G.) and by Canada's National Science and Engineering Research Council (M.F.).

{\em Note added in proof.--} After submitting this manuscript we became aware of a recent preprint by A. Cortijo,\cite{cortijo} wherein an interesting chirality is unearthed in the photon propagator on magnetized topological surfaces. 
The chiral terms in the photon propagator are relativistic and thus negligible in the context of our work, where we employ the dominant non-relativistic piece of the photon propagator. 
The chirality of the interband Coulomb matrix elements in our Eq.~(\ref{eq:coulomb}) stems from the spin-momentum locking of the eigenstates.

\appendix
\begin{widetext}
\section{Numerical Solution of Eq.~(\ref{eq:2body})}

We begin by defining the dimensionless quantities
$\tilde{k}=\hbar v_F k/\Delta$, $\tilde{V}_m(\tilde{k},\tilde{k}')=V_m(\tilde{k},\tilde{k}')/(4\pi \hbar v_F)$, $\tilde{\epsilon}_{\rm ph}=\epsilon_{\rm ph}/(2 \Delta)$ and $\delta\tilde\Sigma=\delta\Sigma/(2\Delta)$, upon which Eq.~(\ref{eq:2body}) can be rewritten as
\begin{equation}
[(\tilde{k}^2+1)^{1/2}+\delta\tilde\Sigma] \chi_m(\tilde{k})-\int_0^\infty d\tilde{k}'\tilde{k}' \tilde{V}_m(\tilde{k},\tilde{k}') \chi_m(\tilde{k}') =\tilde{\epsilon}_{\rm ph}\chi_m(\tilde{k}).\nonumber
\end{equation}
Next, we make a change of variables via $\tilde{k}=\tan(\pi x/2)$ and discretize $x\in[0,1]$ in $N$ segments, 
where the spacing between subsequent $x_i$ and the corresponding weights $w_i$ are chosen according to a $N$-point Gauss-Legendre quadrature method.
This yields\cite{chao1991}
\begin{eqnarray}
\label{eq:disc}
&&[(\tilde{k}_i^2+1)^{1/2}+\delta\tilde\Sigma_i] \chi_m^i-\tilde{V}^{ii}\chi_m^i-\sum_{i\neq j}^N \tilde{V}_m^{i j}\left(\frac{d \tilde{k}}{d x}\right)_j w_j \chi_m^j=\tilde{\epsilon}_{\rm ph}\chi_m^i
\end{eqnarray}
where $\tilde{V}_m^{i j}= \tilde{V}_m(\tilde{k}_i,\tilde{k}_j)$ for $i\neq j$, 
\begin{equation}
\label{eq:vii}
\tilde{V}_m^{ii}=\int_0^\infty d\tilde{k}'\tilde{k}'f(\tilde{k},\tilde{k}')-\sum_{j,j\neq i}^N f(\tilde{k}_i,\tilde{k}_j) \tilde{k}_j \left(\frac{d \tilde{k}}{d x}\right)_j w_j,
\end{equation}
and $f(\tilde{k},\tilde{k}')=2 \tilde{V}_m(\tilde{k},\tilde{k}') \tilde{k}^2 /(\tilde{k}^2+\tilde{k}^{' 2})$.
The first term in Eq.~(\ref{eq:vii}) is evaluated numerically after shifting $\tilde{k}'$ so as to remove the Coulomb singularity.
In the numerical evaluation of Eq.~(\ref{eq:disc}) we take $N=256$; higher values of $N$ do not produce any significant quantitative changes.
\end{widetext}


\begin{thebibliography}{30}
\bibitem{diel} The static dielectric constant is generally smaller in systems with larger bulk bandgaps. 
Arguably, the bulk bandgap in TIs has an upper bound determined by the strength of spin-orbit interactions.
However, other major actors such as phonons might lead to variability in the value of $\epsilon$ even for narrow gap semiconductors.
\bibitem{low energy} Here we limit ourselves to the low-energy effective model. At high energies there is an additional lengthscale, $\sim 1/k_c$, where $k_c$ is the momentum cut-off beyond which the Dirac dispersion is no longer valid.
\end{thebibliography}
\end{document}